\documentclass[aps,preprint,epsfig,rotate]{revtex4}
\begin{document}
\title{On the bound state of the antiproton-deuterium-tritium ion} 

 \author{Alexei M. Frolov}
 \email[E--mail address: ]{afrolov@uwo.ca}

\affiliation{Department of Chemistry\\
 University of Western Ontario, London, Ontario N6H 5B7, Canada}

\date{\today}

\begin{abstract}

The properties of the weakly-bound $S(L = 0)-$state in the $\overline{p}dt$ 
ion are investigated with the use of the results of highly accurate 
computations. The hyperfine structure splitting of this ion is investigated.
We also evaluate the life-time of the $\overline{p}dt$ ion against the nuclear
$(d,t)-$fusion and discuss a possibility to evaluate the corresponding 
annihilation rate(s).

\end{abstract}

\maketitle
\newpage

\section{Introduction}

The boundness of the Coulomb three-body system with unit charges formed 
by the triton $t$ (or tritium nucleus), deuteron $d$ (or deuterium 
nucleus) and one negatively charged antiproton $\overline{p}$ was 
discussed in a number of our earlier studies \cite{FrBi}, \cite{FrTa}. 
It was shown that this system is certainly bound. However, it was also 
found that the $\overline{p}dt$ ion has only one bound (ground) $S(L 
= 0)-$state and this state is, in fact, a weakly-bound state. According 
to the definition of weakly-bound states the ratio of the binding energy 
$\varepsilon$ to the total energy $E$ of each of these states is very 
small, i.e. $\tau = \frac{\varepsilon}{E} \ll 1$. For weakly-bound states 
in Coulomb few-body systems the dimensionless parameter $\tau$ must be 
less than 0.01 (or 1 \%). A number of Coulomb three-body systems (ions) with
unit charges have some weakly-bound states. For instance, the $P^{*}(L = 
1)-$states (excited $P-$states) in the $dd\mu$ and $dt\mu$ ions are 
extremely weakly-bound. However, the ground $S(L = 0)-$states are 
weakly-bound only in a very few such systems. In particular, the ground
$S(L = 0)-$state in the three-body $\overline{p}dt$ ion is weakly-bound. 
It makes the $\overline{p}dt$ ion unique among all Coulomb three-body 
systems with unit charges, since it contains three heavy particles, but has 
very small binding energy.

In general, each weakly-bound state has a number of extraordinary properties. 
Our main goal in this study is to investigate such properties of the ground 
state of the $\overline{p}dt$ ion. First of all, we want to investigate 
the properties which are `typical' for each weakly-bound state in the Coulomb 
three-body systems with unit charges. As follows from the results of numerical
computations of the weakly-bound states many of such properties are 
substantially mass dependent. Even small variations of particle masses produce
noticeable changes in the expectation values of large number of properties. To
avoid this problem one must use extremely accurate variational wave functions.
Briefly, this means that the total and binding energies of such a state must be 
determined to the maximal numerical accuracy which can be achieved in modern 
bound state calculations. To achieve this goal we need to solve the 
corresponding Schr\"{o}dinger equation for bound states in the Coulomb three-body 
system $\overline{p}dt$ to very high numerical accuracy. The highly accurate wave 
function $\Psi$ obtained during this procedure is used later to compute various 
bound state properties of the $\overline{p}dt$ ion. Some of these properties are 
of great interest in numerous applications involving the $\overline{p}dt$ ion.

It is clear that the structure of the bound state(s) in the $\overline{p}dt$ ion 
cannot be explained accurately by ignoring contributions from strong interactions
between antiproton, deuteron and triton \cite{Hu}. Indeed, the actual interparticle 
distances in the ground $S(L = 0)-$state of the $\overline{p}dt$ ion are only in 
$\approx$ 50 times larger than the effective radius of the nucleon-nucleon 
interactions (or $NN-$interactions, for short). Therefore, one can expect that the
strong components of all (three) interparticle potentials in the $\overline{p}dt$ 
ion can change the computed expectation values, i.e. the bound state properties. 
In this study, however, we will ignore all possible contributions from the strong 
components of all interparticle potentials. Our main goal below is to perform 
highly accurate analysis of the Coulomb three-body system $\overline{p}dt$ with 
unit charges. In the next study we are planning to include more realistic 
interaction potentials in our analysis.  

\section{The Hamiltonian and its reduction for weakly-bound systems}

In the non-relativistic approximation the Hamiltonian of the three-body  
$\overline{p}dt$ ion takes the form \cite{LLQ}
\begin{equation}
 H = -\frac{\hbar^{2}}{2 m_{p}} \Bigl( \frac{m_{p}}{m_d} \nabla^{2}_{d}
 + \frac{m_{p}}{m_t} \nabla^{2}_{t} + \nabla^{2}_{\overline{p}} \Bigr) +
 \frac{q_d q_t e^2}{r_{dt}} + \frac{q_d q_{\overline{p}} 
 e^2}{r_{d\overline{p}}} + \frac{q_t q_{\overline{p}} 
 e^2}{r_{t\overline{p}}} \label{Ham}
\end{equation}
where $\nabla_{i} = \Bigl( \frac{\partial}{\partial x_{i}},
\frac{\partial}{\partial y_{i}}, \frac{\partial}{\partial z_{i}} \Bigr)$ and
$i = d, t, \overline{p}$. In Eq.(\ref{Ham}) the notation $\hbar$ stands the reduced 
Planck constant, i.e. $\hbar = \frac{h}{2 \pi}$, and $e$ is the elementary electric 
charge. For the $\overline{p}dt$ ion it is very convenient to perform all bound 
state calculations in proton-atomic units in which $\hbar = 1, m_{\overline{p}} = 
m_p = 1$ and $e = 1$. Here and everywhere below in this study we assume that the 
masses of proton and antiproton exactly equal to each other. The speed of light $c$ 
in the proton-atomic units is $c = \alpha^{-1}$, where $\alpha = \frac{e^2}{\hbar 
c}$ is the fine structure constant. In proton-atomic units the same Hamiltonian, 
Eq.(\ref{Ham}), is written in the form
\begin{equation}
 H = -\frac12 \Bigl( \frac{1}{m_d} \nabla^{2}_{d} + \frac{1}{m_t}
 \nabla^{2}_{t} + \nabla^{2}_{\overline{p}} \Bigr) + \frac{1}{r_{d t}} -
 \frac{1}{r_{d\overline{p}}} - \frac{1}{r_{t\overline{p}}} \label{Ham1}
\end{equation}
where the nuclear masses $m_d$ and $m_t$ of the deuterium and tritium nuclei
must be expressed in terms of the antiproton mass $m_{\overline{p}}$ which 
exactly coincides with the proton mass $m_{p}$. 
 
The highly accurate wave function of the $\overline{p}dt$ ion is obtained during 
numerical solution of the non-relativistic Schr\"{o}dinger equation for Coulomb 
three-body $\overline{p}dt$ system $H \Psi({\bf r}_d, {\bf r}_t, {\bf 
r}_{\overline{p}}) = E \cdot \Psi({\bf r}_d, {\bf r}_t, {\bf r}_{\overline{p}})$, 
where $H$ is from Eq.(\ref{Ham1}) and $E < 0$ is the total energy of this ion. Then 
by using this highly accurate wave function we can compute a number of different 
expectation values. These expectation values are considered as the bound state 
properties of the ground (bound) state in the $\overline{p}dt$ ion.

The Hamiltonian of the weakly-bound Coulomb three-body systems, Eq.(\ref{Ham1}), 
can be reduced to the sum of the two separated Hamiltonians, i.e. $H = H_i + H_o$, 
where the Hamiltonian $H_i$ of the central (or compact) sub-system $t\overline{p}$ 
and $H_o$ is the Hamiltonian of the outermost particle. For the $\overline{p}dt$ 
ion these two Hamiltonians are (in proton-atomic units)
\begin{equation}
 H_i = -\frac12 \Bigl( \frac{1}{m_t} \nabla^{2}_{t} + \nabla^{2}_{\overline{p}} 
 \Bigr) - \frac{1}{r_{t\overline{p}}} \label{Ham11}
\end{equation}
and 
\begin{equation}
 H_o = -\frac{1}{2 m_d} \nabla^{2}_{d}  + \frac{1}{r_{d t}} -
 \frac{1}{r_{d\overline{p}}} \label{Ham2}
\end{equation}
The interaction potential $\frac{1}{r_{d t}} - \frac{1}{r_{d \overline{p}}}$ in the 
Hamiltonian $H_o$, Eq.(\ref{Ham2}), is the short-range potential, which is, in fact, 
a polarization potential. This polarization potential acts between the positively 
charged deuterium nucleus (electric charge equals +1) and neutral $t\overline{p}$ 
quasi-atom. In this two-body picture the deuterium nucleus produces `sufficiently 
large' polarization of the $t\overline{p}$ quasi-atom. The interaction between the 
$d^{+}$ ion and polarized $t\overline{p}$ quasi-atom leads to the formation of the 
weakly-bound state in the $\overline{p}dt$ ion. The boundness of such an ion can
accurately be evaluated with the use of perturbation theory (see below). Note also 
that the actual, three-particle potential $\frac{1}{r_{d t}} - \frac{1}{r_{d 
\overline{p}}}$ in the $\overline{p}dt$ ion is the difference between the attractive 
and repulsive Coulomb two-body potentials. All terms in such a potential which have
non-zero long range asymptotics are compensate each other. Therefore, at large distances 
between the central cluster $\overline{p}t$ and deuterium nucleus $d^{+}$ such a 
potential almost equals to zero. This allows one to reduce the original, very complex, 
three-body problem to the interaction of the central, two-body cluster and one deuterium 
nucleus, i.e. to the two-body problem. Briefly, we can say that the deuterium nucleus 
produces electric polarization of the central cluster. If such a polarization is 
relatively large, then one (or even a few) bound state arise which are stable against 
dissociation.  

The Schr\"{o}dinger equation $H \Psi = E \cdot \Psi$ with the Hamiltonian given by 
Eq.(\ref{Ham11}) is solved exactly and gives us the hydrogen-like wave functions 
$\psi({\bf r}_{t\overline{p}})$ of the two-body quasi-atom $t\overline{p}$ (or 
$\overline{p}t$). Then, by using these hydrogenic wave functions we can compute the 
expectation value of the $H_o$ operator, Eq.(\ref{Ham2}), which is an operator in 
respect to the (Cartesian) coordinates of the deuterium nucleus $d$. The origin of 
the Cartesian coordinates of the deuterium nucleus coincides with the center of mass 
of the $t\overline{p}$ quasi-atom. The kinetic energy of the deuterium nucleus in 
Eq.(\ref{Ham2}) does not change, if we assume that all wave functions $\psi({\bf 
r}_{t\overline{p}})$ (bound state wave functions) have the unit norm. The expectation 
value of the sum of the two Coulomb potentials in Eq.(\ref{Ham2}) can be replaced by 
some effective potential $V({\bf r}_d)$, which in the first order approximations is the 
central potnetial, i.e. $V({\bf r}_d) = V(r_d)$. The explicit form of this effective 
potential can be obtained with the use of the perturbation theory. This problem is 
considered in the next Section. It is clear that such a picture is only an approximation, 
since we have ignored a number of additional factors, e.g., contribution from the 
hydrogenic wave functions of continuous spectrum. Nevertheless, the overall accuracy of 
this model is surprisingly high and it still often used for many three-body systems with 
unit charges.

\section{Polarisation potential}

As shown above, by using the short-range polarisation potential in Eq.(\ref{Ham2}) one can 
reduce analysis of the original three-body Coulomb system with unit charges to the two 
`equivalent' two-body problems, Eqs.(\ref{Ham11}) - (\ref{Ham2}). In reality, such an 
`equivalent' replacement is very difficult to complete, since there are many relations 
between the explicit form(s) of the model two-body potential in Eq.(\ref{Ham2}) and its 
properties which can be observed in actual, three-body experiments. For instance, as follows 
from the general theory of the two-body bound states in the non-relativistic potential field 
one finds for the total number $N(\ell)$ of bound states with the angular momentum $\ell$
\begin{equation}
 N(\ell) \leq \frac{2 m}{(2 \ell + 1) \hbar^2} \int_0^{\infty} r \mid V(r) \mid dr 
 \label{Barg}
\end{equation} 
where the central potential $V(r)$ is the polarisation potential mentioned above. For the 
$\overline{p}dt$ ion we have $\ell = 0$ and the mass $m$ in the Bargmann inequality, 
Eq.(\ref{Barg}), (see, e.g., \cite{Schmidt} and references therein) is the mass of the 
deuterium nucleus. A slightly more complicated analysis indicates that the mass $m$ in 
Eq.(\ref{Barg}) must be chosen as the reduced mass of the deuterium $d$ and $t\overline{p}$ 
quasi-atom. Finally, for the $\overline{p}dt$ ion one finds from Eq.(\ref{Barg}) (in 
proton-atomic units)
\begin{equation}
 N(\ell=0) \leq 2 \cdot \frac{m_{d} (m_{t} + m_{\overline{p}})}{m_{d} + m_{t} + 
 m_{\overline{p}}} \int_0^{\infty} r \mid V(r) \mid dr \label{B1}
\end{equation} 
Note that all particle masses in the $\overline{p}dt$ ion are comparable to each other.
Therefore, the distribution of the electric charge density in this ion is spherically 
symmetric (or almost spherically symmetric). The first order correction upon the 
charge-dipole interaction between the deuterium nucleus $d^{+}$ and $t\overline{p}$ system 
is zero, since the spatial distribution of electric charge in the central $t\overline{p}$ 
cluster is spherically symmetric. The second order perturbation theory leads to the 
following general form for the interaction $V(r)$ potential between the deuterium nucleus 
$d^{+}$ and neutral quasi-atom $t\overline{p}$ in the $\overline{p}dt$ ion
\begin{equation}
 V(r) = \frac{A}{(r + a)^4} + \frac{B}{(r + b)^5} \label{eq7}
\end{equation} 
The four constants $A,a, B$ and $b$ in this potential must be in agreement with the 
prediction which follows from Eq.(\ref{B1}) and with other experimental data obtained, e.g., 
from the scattering of the deuterium nucleus at the $\overline{p}t$ system (at different 
energies). In addition, the two-body system with the interaction potential $V(r)$ must have 
the same (or almost the same) bound state properties as the $\overline{p}dt$ ion. 

It should be mentioned that the two-body polarisation potential $V(r)$, Eq.(\ref{eq7}), 
carefully reconstructed with the use of the known scattering data and bound state 
properties of the $\overline{p}dt$ ion  is of some interest in a large number of applications. 
For instance, the total number of bound states and their approximate geometrical and 
dynamical properties can be obtained with the use of such a potential. On the other hand, 
such a potential is only approximation to the actual three-body potential. Therefore, it 
is hard to expect that all bound state properties of the $\overline{p}dt$ ion determined 
with the model two-body potential will be in good numerical agreement with the actual 
properties. 

\section{Variational calculations}

In general, the highly accurate computations of weakly-bound states in Coulomb three-body 
systems with unit charges are difficult to perform. The main reason for this is obvious, 
since all traditional variational expansions contains `pieces' which describe 
contributions of fragments from the unbound spectra of two-body systems. In actual 
computations this leads to very slow convergence rate at large dimensions, i.e. when large 
number(s) of basis functions $N$ are used in computations. Traditionally, the highly 
accurate variational computations of bound states in Coulomb three-body systems with unit 
charges are performed with the use of the exponential variational expansion in 
perimetric/relative coordinates. The explicit form of such an expansion for $S(L = 
0)$-states is
\begin{eqnarray}
 \Psi = \sum_{i=1}^{N}  C_{i} \phi_i(r_{32},r_{31},r_{21}) 
 \exp(-\alpha_{i} u_1 - \beta_{i} u_2 - \gamma_{i} u_3) \exp(\imath \delta_{i} u_1 + 
 \imath e_{i} u_2 + \imath f_{i} u_3) \label{exp} 
\end{eqnarray}
where $C_{i}$ are the linear (or variational) parameters, $\alpha_i, \beta_i, \gamma_i, 
\delta_i, e_i$ and $f_{i}$ are the non-linear parameters and $\imath$ is the imaginary 
unit. the function $\phi_i(r_{32},r_{31},r_{21})$ is the polynomial (usually quaratic)
function of the three relative coordinates $r_{ij}$. The coefficients of this function
are fixed and never varied in calculations. The notations $u_1, u_2$ and $u_3$ in 
Eq.(\ref{exp}) are the three perimetric coordinates: $u_i = \frac12 (r_{ij} + r_{ik} -
r_{jk})$. It can be shown that three periemetic coordinates are independent of each other
and each of them varies between 0 and $+\infty$. 

The variational expansion, Eq.(\ref{exp}), provides extremely high accuracy in the bound 
state computations of arbitrary three-body systems (for more discussions, see, e.g., 
\cite{Fro2001}). For highly accurate calculations of the ground state in the 
$\overline{p}dt$ ion we can assume that all non-linear parameters $\delta_i, e_i, f_i$ 
equal zero idential, i.e.  $\delta_i = 0, e_i = 0, f_i = 0$ for $i = 1, \ldots, N$. This
means that all varied non-linear parameters in the trial wave function, Eq.(\ref{exp}), 
are real. This allows one to re-write the formula, Eq.(\ref{exp}), to the form
\begin{eqnarray}
 \Psi = \sum_{i=1}^{N}  C_{i} \exp(-\alpha_{i} r_{31} - \beta_{i} r_{31} 
  - \gamma_{i} r_{21})  \label{exp1} 
\end{eqnarray}
which is called the exponential variational expansion in the relative coordinates $r_{32},
r_{31}$ and $r_{21}$. In general, the total energy of the ground state of the 
$\overline{p}dt$ ion uniformly depends upon the total number of basis functions $N$,
Eq.(\ref{exp1}), used in calculations. 

Now, the phenomenon of slow convergence at large dimensions can be illustrated with the 
use of the asymptotic formula for the $E(N)$ dependence
\begin{equation}
 E(N) = E(\infty) + \frac{A_1}{N^{\delta}} + \frac{A_2}{N^{\delta + 1}} \label{asy}
\end{equation} 
at large $N$. The formula, Eq.(\ref{asy}), includes the four parameters $E(\infty), A_1, 
A_2$ and $\delta$. The asymptotic energy $E(\infty)$ is the improved total energy which 
is closer to the `exact' answer than the computed $E(N_i)$ values. For bound states 
in the Coulomb three-body systems with unit charges, which are not weakly-bound, the 
parameter $\delta$ in Eq.(\ref{asy}) does not change noticeably with $N$. For weakly-bound 
states in such systems the parameter $\delta$ at large $N$ becomes $N$-dependent. In many 
actual cases it decreases when $N$ increases.
 
To perform numerical calculations in this work we have used another approach devaloped in
our papers \cite{Fro2001}. In that approach all actual non-linear parameters of the method 
(usually 28 - 40 non-linear parameters) are carefully optimized at some relatively large 
dimension, e.g., for $N$ = 1800 and for $N$ = 2200. Then the total number of basis functions 
has been increased to $N$ = 3500 - 3840. The optimized values of the non-linear parameters 
have not been changed during this (last) step. This simple method allows one to obtain 
results which are significantly more accurate than it is possible to achieve by using 
old-fashion numerical procedures with the same numbers of basis functions. Finally, by using 
our highly accurate wave functions we can determine the expectation values of various bound 
state properties. This problem is discussed in the next Section.  

\section{Bound state properties}

The total energy and some other bound state properties of the $\overline{p}dt$ ion can be 
found in Tables I and II. Table I contains the total bound state energies $E$ and the expectation
values of all three inter-particle delta-functions $\langle \delta({\bf r}_{12}) \rangle, \langle 
\delta({\bf r}_{13}) \rangle, \langle \delta({\bf r}_{23}) \rangle$ determined in proton-atomic 
units $\hbar = 1, e = 1$ and $m_{p} = 1$. Table II includes some other expectation values of the 
bound state properties of this ion. These bound state properties are determined as the expectation 
values of the corresponding operators, e.g. for the operator $\hat{X}$ we write its expectation 
value $X$:
\begin{equation}
 X = \frac{\langle \Psi \mid \hat{X} \mid \Psi \rangle}{\langle \Psi 
 \mid \Psi \rangle}
\end{equation}
where $\hat{X} = r^n_{ij}$, where $n$ = -2, 1, 2, 3, 4 , $\hat{X} = \delta({\bf r}_{ij}), \hat{X} 
= \delta({\bf r}_{ij}) \delta({\bf r}_{ik})$, etc. Here and below we use the cyclic notations 
$(i, j, k) = (1, 2, 3)$ which are very convenient for three-body systems. Note that all 
distances in the $\overline{p}dt$ ion are in $M_p \approx 1836$ times shorter than the 
electron-nuclear distances in regular hydrogen atom and hydrogen ion. This means that the 
distance between the antiproton and tritium/deuterium nuclei in the ground state of the 
$\overline{p}dt$ ion are $\approx 2.5 - 2.6 \cdot 10^{-12}$ $cm$. Briefly, these distances are 
only in $\approx$ 20 - 30 times larger than the corresponding nucleon-nucleon distances in the light 
few-nucleon nuclei, e.g., in the ${}^3$H, ${}^3$He and ${}^4$He nuclei. This indicates clearly that 
strong interactions between deuteron, triton and antiproton can contribute significantly to the 
structure and properties of the bound states in the $\overline{p}dt$ ion. In particular, such a 
contribution will be large for all properties which include the interparticle delta-functions, i.e. 
$\delta({\bf r}_{ij})$ values.      

The quality of the computed delta-functions $\delta({\bf r}_{ij})$ can be checked by comparing the
computed and predicted values of the corresponding cusp-values $\nu_{ij}$ which are defined by the
following relations
\begin{equation}
 \nu_{ij} = \frac{\langle \Psi \mid \delta({\bf r}_{ij}) \frac{\partial}{\partial r_{ij}} 
 \mid \Psi \rangle}{\langle \Psi \mid \delta({\bf r}_{ij}) \mid \Psi \rangle}
\end{equation}
where $(ij) = (ji)$ = (12), (13), (23). For few-body systems interacting by the Coulomb and 
Yukawa-type forces the numerical values of $\nu_{ij}$ can be predicted and they are not equal 
zero identically. In particular, for the Coulomb three-body system one finds:
\begin{equation}
 \nu_{ij} = q_i q_j \frac{m_i m_j}{m_i + m_j} 
\end{equation}
where $q_i$ and $q_j$ are the electric charges and $m_i$ and $m_j$ are the masses of the particles.

By using any two of the three interprticle delta-functions we can define the three-particle 
delta-function $\delta_{123} = \delta({\bf r}_{ij}) \cdot \delta({\bf r}_{ik})$. The $\langle 
\delta_{123} \rangle$ expectation value equals to the probability to detect all three particles at 
one spatial `non-relativistic' point, i.e. inside of the volume $V = \alpha^3 a^3_0$, where $\alpha
= \frac{e^2}{\hbar c}$ is the fine-structure constant, while $a_0 = \frac{\hbar^2}{m_e e^2}$ is the 
Bohr radius. This expectation value plays an important role for some systems, e.g., it is used to 
predict the one-photon annihilation in the Ps$^{-}$ ion. Formally, by using the three-particle 
delta-function $\delta_{123}$ one can try to construct the three-particle cusp operator $\nu_{123}$, 
which equals to the product of the three-particle delta-function $\delta_{123}$ and a second order 
(partial) derivative $\frac{\partial^2}{\partial r_{ij} \partial_{ik}}$. However, as follows from 
the general theory \cite{Fock} the expectation value of such an operator is infinite for an arbitrary 
Coulomb three-body system. 

The expectation values of the spatial momenta $r^{m}_{ij}$ are defined as follows:
\begin{equation}
 \langle r^m_{ij} \rangle = 
 \int \int \int \Psi(r_{12}, r_{13}, r_{23}) r_{ij}^m \Psi(r_{12}, r_{13}, r_{23}) 
 r_{12} r_{13} r_{23} dr_{12} dr_{13} dr_{23}
\end{equation}
This integral is reduced to the form of the one-dimensional integral of the product of the 
one-particle density matrix and $r^m_{ij}$. To simplify the following transformations below we 
shall restrict ourselves to the consideration of the bound $S(L = 0)-$states only. In this case, the 
wave function $\Psi$ is the function of the three scalar interparticle coordinates $r_{12}, r_{13}$ 
and $r_{23}$ only, i.e. $\Psi = \Psi(r_{12}, r_{13}, r_{23})$. Therefore, we can define the 
one-particle density matrix 
\begin{equation}
 \rho(r_{12}) = \int \int \Psi(r_{12}, r_{13}, r_{23}) \Psi(r_{12}, r_{13}, r_{23}) 
 r_{13} r_{23} dr_{13} dr_{23}
\end{equation}
and write the following formula for the $\langle r^m_{12} \rangle$ expectation value 
\begin{equation}
 \langle r^m_{12} \rangle = \int_0^{+\infty} r_{12}^{m+1} \rho(r_{12}) dr_{12} 
\end{equation}
The idea of this method of `spatial moments' developed in the middle of 1960's was simple and 
transparent. Indeed, if we can determine a sufficient number of `spatial moments', then we can 
reconstruct, in principle, all three one-particle density matrixes $\rho(r_{12}), \rho(r_{13})$ and 
$\rho(r_{23})$. In turn, this will lead us to useful conclusions about spatial distributions of the 
two and three particles in the system. However, this idea has never been applied to reconstruct the 
actual density matrixes. Furthermore, since the middle of 1960's we can approximate the wave 
function of any three-body system to very high numerical accuracy. By using these wave functions we 
can determine all spatial momenta to extremely high accuracy (see Table II). However, a few 
questions about spatial momenta are still remain unaswered for three-body systems. The most 
interesting question is related to the `singular momenta', i.e. to the $\langle r^m_{ij} \rangle$ 
expectation values with $m = -3, -4, -5, \ldots$. These expectation values are singular, but their 
principal and regular parts are needed in some problems, e.g., the values $\langle r^{-3}_{ij} 
\rangle$ are used to evaluate the lowest-order QED correction (or Araki-Sucher correction) for Coulomb 
two-electron atoms and ions (see, e.g., \cite{Fro07} and references therein). Our Table II contains
the epectation values of the regular part of the $\langle r^{-3}_{ij} \rangle$ expectation value which 
is designated as $\langle r^{-3}_{ij} \rangle_R$. These expectation values are related to the 
corresponding principal parts ($\langle r^{-3}_{ij} \rangle$) by the following relations (see, e.g., 
\cite{Fro07} and references therein) $\langle r^{-3}_{ij} \rangle = \langle r^{-3}_{ij} \rangle_R + 
4 \pi \langle \delta({\bf r}_{ij}) \rangle$.   

Table II also contains expectation values which are used to measure various interparticle correlations 
in three-body systems. The spatial interparticle correlations are evaluated with the use of the three
$cosine$-functions $\tau$:
\begin{equation}
 \tau_{jk} = \langle \frac{{\bf r}_{ij} \cdot {\bf r}_{ik}}{r_{ij} r_{ik}} \rangle  
\end{equation}
where ${\bf a} \cdot {\bf b}$ means the scalar product of the two vectors (${\bf a}$ and ${\bf b}$), 
$r_{ij} = r_{ji}$, but ${\bf r}_{ij} = -{\bf r}_{ji}$, since ${\bf r}_{ij} = {\bf r}_{i} - {\bf r}_{j}$. 
The sum of the three $\tau_{jk} $ values always exceeds unity, i.e. it can be written in the form 
\begin{equation}
 \tau_{12} + \tau_{13} + \tau_{23} = 1 + 4 \cdot f   
\end{equation}
where the $f-$value is  
\begin{eqnarray}
 f = \langle \Psi \mid \frac{u_1 u_2 u_3}{r_{12} r_{13} r_{23}} \mid \Psi \rangle = \int \int \int    
 \mid \Psi(r_{12}, r_{13}, r_{23}) \mid^2 \frac{u_1 u_2 u_3}{r_{12} r_{13} r_{23}} r_{12} r_{13} 
 r_{23} dr_{12} dr_{13} dr_{23} \nonumber \\
 = 2 \int \int \int \mid \Psi(u_1, u_2, u_3) \mid^2 u_1 u_2 u_3 du_1 du_2 du_3 
\end{eqnarray}

In some earlier works the expectation values of the three scalar products ${\bf r}_{ij} \cdot {\bf 
r}_{ik}$ were proposed to be used to describe spatial interparticle correlations in the three-body
systems. Note that the three interparticle vectors ${\bf r}_{ij}, {\bf r}_{ik}, {\bf r}_{kj}$ always 
form a triangle. Therefore, we can write 
\begin{equation}
    {\bf r}_{ki} + {\bf r}_{ij} = {\bf r}_{kj} \; \; \; , \; or \; \; \;
    {\bf r}_{ik} \cdot {\bf r}_{ij} = \frac12 (r^2_{ik} + r^2_{ij} - r^2_{kj})
\end{equation}   
In other words, the expectation values of all three scalar products ${\bf r}_{ij} \cdot {\bf 
r}_{ik}$ can be expressed from the known $\langle r^2_{ij} \rangle$ values. This means that these 
scalar products are not independent properties. Analogous situation can be found for the three scalar 
products ${\bf p}_{i} \cdot {\bf p}_{j}$ of each two single particle momenta. It follows from the 
conservarion of the total momentum ${\bf P}$ and its three components $({\bf P})_i$. Indeed, in the 
center-of-mass system we can write ${\bf P} = {\bf p}_1 + {\bf p}_2 + {\bf p}_3 = 0$. From here one 
finds three following identities 
\begin{equation}
    {\bf p}_{i} \cdot {\bf p}_{j} = \frac12 (p^2_{i} + p^2_{j} - p^2_{k})
\end{equation}   
The same identities can be written for the corresponding expectation values. Therefore, the three 
expectation values of the scalar products ${\bf p}_{i} \cdot {\bf p}_{j}$ are uniformly related to
the single particle `kinetic energies' $\frac12 \langle p^2_i \rangle$ and cannot be used to describe 
dynamical correlations between three point particles as independent values. 

\section{Annihilation and fusion rates}
 
In reality, the three-body $\overline{p}dt$ ion is not stable, since it decays either by the 
proton-antiproton annihilation (at rest), or by the nuclear $(d,t)-$fusion. Note that the 
proton-antiproton annihilation in the $\overline{p}dt$ ion proceeds as the tritium-antiproton 
annihilation, or deuterium-antiproton annihilation. In general, the proton-antiproton (or 
triton-antiproton and deuteron-antiproton) annihilation can proceed with the use of many dozens 
reaction channels (see, e.g., \cite{Barel} and references therein), e.g.,
\begin{eqnarray}
 \overline{p} d t \rightarrow \pi^{0} + \pi^{+} + \pi^{-} + d + 2 n + Q_1 (MeV) \label{annih} \\
 \overline{p} d t \rightarrow \pi^{\pm} + K^{0} + t + n + Q_2 (MeV) \label{annih2} 
\end{eqnarray}
where $\pi^{0}, \pi^{\pm}$ and $K^{0}$ are the neutral and charged pions and kaons, respectively.
The energy released during antiproton annihilation is usually around hundreds and even thousands of 
$MeV$ \cite{Barel} (this depends upon the annihilation channel). 

The reaction of nuclear fusion takes the form 
\begin{eqnarray}
   \overline{p} d t \rightarrow \overline{p} + {}^4{\rm He} + n + 17.1 MeV
\end{eqnarray}
where the antiproton $\overline{p}$ plays the role of a particle-`catalyzator' which simply brings the 
two positive particles closer to each other. Formally, the distance between the nuclei of deuterium and 
tritium is in $m_p \approx 1836.15$ times shorter than in the DT$^{+}$ molecular ion and in $\approx$ 8.880 
times shorter than in the ground state of the muonic molecular ion $dt\mu$. This allows one to evaluate the
rate of the (d,t)-nuclear fusion. The approximate value is $\approx 1 \cdot 10^{13} - 2.5 \cdot 10^{13}$ 
$sec^{-1}$. The inverse value gives the approximate life-time $\tau_f$ of the $\overline{p}dt$ ion against 
nuclear $(d,t)-$fusion ($\tau_f \approx 4 \cdot 10^{-14} - 1 \cdot 10^{-13}$ $sec$). This result is based on 
the quasi-classical approximation which provides a reasonable accuracy for muonic molecular ions, but can be 
very approximate for the $\overline{p}dt$ ion. Note also that the probability of formation of the bound, 
two-body $\overline{p} {}^{4}$He system (also called $\overline{p}\alpha$-system) after such a reaction 
(catalyzator poisoning) in the $\overline{p}dt$ ion in is relatively large ($\approx$ 5 \% according to our 
approximate evaluations). 

Analogous evaluation of the life-time of the $\overline{p}dt$ ion against antiproton annihilation is 
significantly more complicated. Currently, the both tritium-antiproton and deuterium-antiproton annihilations 
at small and very small energies of the colliding particles are not well studied phenomena. In particular, 
there are some deviations between experimental data and theoretical predictions based on different theoretical 
models (see, e.g., \cite{2009}, \cite{1997}, \cite{1999} and references therein). However, the main problem for 
low-energy analysis is related with the fact that currently there is no experimental data below 38 $MeV/c$ (the 
incident moment of the antiproton). Very likely, at small energies there are large differences between 
annihilation rates obtained for the triplet and singlet states of proton-antiproton pairs. Moreover, the role 
of antiproton annihilation at neutron(s) at small energies can be evaluated only approximately (see discussion 
below). It makes almost impossible any realisic evaluation of the tritium-antiproton and deuterium-antiproton 
annihilation rates at small/zero energies of the colliding particles. 

Nevertheless, a few predictions for the ground state in the $\overline{p}dt$ ion can be made. Indeed, by 
using our expectation values of the interparticle delta-functions determined for the ground state of the 
$\overline{p}dt$ ion (see Table I) we can evaluate the relative contributions of the antiproton annihilation 
at the triton and deuteron nuclei, respectively. Indeed, for the antiproton annihilation rates ($\Gamma$) at 
the deuteron and/or triton we can write the following formulas:
\begin{equation}
 \Gamma_{\overline{p}d} = (a_{\overline{p}p} + a_{\overline{p}n}) \langle \delta_{\overline{p}d} 
 \rangle \; \; \; , \; \; \; \Gamma_{\overline{p}t} = (a_{\overline{p}p} + 2 a_{\overline{p}n}) 
 \langle \delta_{\overline{p}t} \rangle  
\end{equation}
where the notations $p, \overline{p}, n$ stand for the proton, antiproton and neutron, respectively. From here 
one finds 
\begin{equation}
 \frac{\Gamma_{\overline{p}t}}{\Gamma_{\overline{p}d}} = \Bigl( \frac{2 \xi + 1}{\xi + 1} \Bigr)
 \frac{\langle \delta_{\overline{p}t} \rangle}{\langle \delta_{\overline{p}d} \rangle}
 \approx 7.816177 \cdot \Bigl( \frac{\xi + \frac12}{\xi + 1} \Bigr) \label{factor}
\end{equation}
where $\xi = \frac{a_{\overline{p}n}}{a_{\overline{p}p}}$ is the ratio of the corresponding `basic'
annihilation rates (the rates of the neutron-antiproton and proton-antiproton annihilation at zero interparticle 
distances). As follows from the results of numerous experiments with heavy antiprotonic atoms (\cite{2001} and 
references therein) the ratio $x$ is close to unity. If $\xi = 1$, then from Eq.(\ref{factor}) one finds that for 
the ground state in the $\overline{p}dt$ ion $\frac{\Gamma_{\overline{p}t}}{\Gamma_{\overline{p}d}} = 5.86213$. 
It is extremely interesting to evaluate the actual ratio $\xi$ for the $\overline{p}dt$ ion by performing direct 
experiments. This will give us an improtant information about the antiproton-neutron annihilation at rest. Also, 
it follows from Table I for the $\overline{p}dt$ ion the antiproton annihilation rate at the triton is $\approx$ 
5.86213 times more likely than annihilation at the deuteron (if $\xi = 1$). It is clear that experiments with the 
$\overline{p}dt$ ion can be used for accurate measurments of the $\xi$ ratio at small/zero energies of the
colliding particles.

\section{Hyperfine structure}

The hyperfine structure of the $\overline{p}dt$ ion is of great experimental interest, since 
the magnetic moments of the proton and antiproton are equal to each other by their absolute 
values, but have opposite sings. Note also that all three particles in the  $\overline{p}dt$ 
ion have non-zero magnetic moments (or spins). Therefore, we can expect that the hyperfine
structure of the ground state of the $\overline{p}dt$ ion will be sufficiently rich. Moreover,
such a structure and corresponding hyperfine structure splittings can easily be observed in
modern experiments. However, the determined values of the hyperfine splittings of the 
$\overline{p}dt$ allows us to understand a large number of interesting details related to the 
long-range asymptotics of the strong interparticle interactions.

For the bound $S(L = 0)-$state in the $\overline{p}dt$ ion the hyperfine structure is 
determined by solving the eigenvalue problem for the Hamiltonian $(\Delta H)_{h.s.}$ which is 
responsible for the spin-spin (or hyperfine) interactions. The general formula for such a 
hyperfine Hamiltonian $(\Delta H)_{h.s.}$ in the case of the $\overline{p}dt$ ion is written 
as the sum of the three following terms. Each of these terms is proportional to the product of 
the factor $\frac{2 \pi}{3} \alpha^2$ and the expectation value of the corresponding 
(interparticle) delta-function. The third (additional) factor contains the corresponding 
$g-$factors (or gyromagnetic ratios) and scalar product of the two spin vectors. For instance, 
for the $pd\mu$ ion this formula takes the form (in atomic units $\hbar = 1, e = 1, m_e = 1$) 
(see, e.g., \cite{LLQ})
\begin{eqnarray}
  (\Delta H)_{h.s.} = \frac{2 \pi}{3} \alpha^2 \frac{g_{\overline{p}} g_d}{m^2_p}
  \langle \delta({\bf r}_{\overline{p}d}) \rangle ({\bf s}_{\overline{p}} \cdot {\bf s}_d) +
  \frac{2 \pi}{3} \alpha^2 \frac{g_{\overline{p}} g_t}{m^2_p}
  \langle \delta({\bf r}_{\overline{p}t}) \rangle ({\bf s}_{\overline{p}} \cdot {\bf s}_t) 
 \nonumber \\
 + \frac{2 \pi}{3} \alpha^2 \frac{g_d g_t}{m^2_p}
   \langle \delta({\bf r}_{dt}) \rangle ({\bf s}_d \cdot {\bf s}_t)
   \label{spl1}
\end{eqnarray}
where $\alpha = \frac{e^2}{\hbar c}$ is the fine structure constant and $m_p = m_{\overline{p}}$ 
is the proton mass. Note also that the presence of the common factor $\frac{1}{m^2_p}$ in all
three terms in Eq.(\ref{spl1}) indicates clearly that the proton-atomic units mentioned
above are more appropriate for this problem. Nevertheless, below we shall continue to use the 
usual atomic units $\hbar = 1, e = 1$ and $m_e = 1$. To recalculate the expectation values of the 
$\delta-$functions from proton-atomic to atomic units one needs to use the factor $m_p^3 \approx 
6.1905094020072 \cdot 10^9$. The expression for $(\Delta H)_{h.s.}$, Eq.(\ref{spl1}), is, in 
fact, an operator in the total spin space which has the dimension $(2 s_p + 1) (2 s_d + 1) (2 s_t 
+ 1) = 12$. In our calculations we have used the following numerical values for the constants and 
factors in Eq.(\ref{spl1}): $\alpha = 7.297352586 \cdot 10^{-3}$ and $m_{\overline{p}} = 
1836.1527012 m_e$. The $g-$factors for the antiproton, deuteron and triton are deteremined from 
the formulas: $g_p = \frac{{\cal M}_{pi}}{I_p}, g_d = \frac{{\cal M}_d}{I_d}$ and $g_t = 
\frac{{\cal M}_t}{I_t}$, where ${\cal M}_a$ and $I_a$ are the corresponding magnetic moments and 
spins of the particle $a$. For the triton, deuteron and antiproton we have $I_t = \frac12, I_d = 
1$ and $I_p = \frac12$, while the magnetic moments of these particles (in nuclear magnetons) are 
$M_{\overline{p}} = -2.792847386, {\cal M}_d = 0.857438230$ and ${\cal M}_t = 2.97896247745$.

The hyperfine structure and hyperfine structure splittings of the ground state of the 
$\overline{p}dt$ ion can be found in Table III. Traditionally, these values are expressed in Mega
Hertz, or $MHz$. To-recalculate our results from atomic units to $MHz$ we used the conversion 
factor 6.57968392061 $\cdot 10^9$ $MHz/a.u.$ \cite{CRC}. As follows from Table III the hyperfine
structure of the $\overline{p}dt$ ion include twelve levels which are separated into the
four following groups: (1) the group of five spin states with $J = 2$, (2) the upper group of 
three states with $J = 1$, (3) one state with $J = 0$ and (4) the lower group of three states 
with $J = 1$. The energy differences between these four groups of states are: 1.112524269$\cdot 
10^8$ $MHz$, 6.853155051$\cdot 10^9$ $MHz$ and 5.977402022$\cdot 10^7$ $MHz$. These values can be 
measured quite accurately by using modern experimental technique developed to measure atomic 
hyperfine splittings.   

Note again that such a hyperfine structure corresponds to the three-body $\overline{p}dt$ ion with 
the pure Coulomb interaction potentials between particles. In reality, corrections to the strong 
interactions between particles will change the expectation values of all interparticle 
delta-functions. Therefore, we can expect some changes in the hyperfine structure and hyperfine 
structure splittings. Such corrections to the hypefine structure splitting of the $\overline{p}dt$
ion will be the goal of our next study.
 
\section{Conclusion}

We have determined the total energy and some bound state properties of the weakly-bound (ground) $S(L 
= 0)-$state in the $\overline{p}dt$ ion. It is shown that this weakly-bound state has a `two-body' 
cluster structure, which is represented as the motion of the deuterium nucleus $d^{+}$ in the central 
field of the `central' $\overline{p}t$ cluster. The interaction between the deuterium nucleus $d^{+}$ 
and the two-body $\overline{p}t$ cluster is the regular dipole-charge interaction, which is often 
called and considered as the polarization potential. The use of the two-body cluster model and
approximate reconstruction of the polarization potential allows one to evaluate qualitatively many
bound state properties of the $\overline{p}dt$ ion. In highly accurate calculations of the ground 
state of the $\overline{p}dt$ ion performed for this study we have used our advanced optimization
procedure developed in \cite{Fro2001} and later improved for large dimensions. This strategy works 
perfectly for all Coulomb three-body systems, including weakly-bound and cluster systems.

A number of bound state properties of the $\overline{p}dt$ ion have been determined to very high
numerical accuracy with the use of our bound state wave functions. In particular, by using the
expectation values of the delta-functions we have investigated the hyperfine structure of the ground 
$S(L = 0)-$state in the $\overline{p}dt$ ion. In this ion there are twelve levels of hyperfine
structure which are separated into four different groups. The energy differences between these groups 
of states are $\Delta_{12}$ = 1.112524269$\cdot 10^8$ $MHz$, $\Delta_{23}$ = 6.853155051$\cdot 10^9$ 
$MHz$ and $\Delta_{34}$ = 5.977402022$\cdot 10^7$ $MHz$. These values are called the hyperfine
structure splittings. The obtained values of the hyperfine structure splittings must be compared
with the corresponding experimental values.   

Another interesting question for this ion is related to a direct comparison of the rates of the $(d,
t)-$nuclear fusion and rates of the antiproton annihilation at the deuteron and triton, 
respectively. These problems have solutions which are very sensitive to the contribution from 
strong interparticle interactions. Therefore, the comparison of the future experimental results and 
theoretical predictions will lead us to a large number of improtant conclusions about long-range
asymptotics of the strong interactions. Currently, we have developed a number of effective 
numerical methods for such computations. Moreover, as follows from the results of fisrt 
calculations performed with the use of `realistic' interparticle potentials (Coulomb potential
plus term which describes strong interactions) we can say that the contributions from the 
antiproton-triton and antiproton-deuteron strong interactions play the leading role. Analogous  
contribution from the deuteron-triton realistic potential is significantly smaller. Therefore, to
make accurate predictions for the ground state of the $\overline{p}dt$ ion we need to know the 
interaction potential between the antiproton and triton and deuteron, respectively. The contribution 
from the triton-deuteron strong interaction is also improtant, but it is not crucial. In general, the 
accurate reconstruction of the potential of strong interactions require a detailed knowledge of all 
its scalar, spin-orbital (${\bf L} \cdot {\bf S}$) and tensor components. To reconstruct such a 
potential for each pair of interacting particles one needs to know a large amount of scattering data 
\cite{Babik} and additional information about all bound two-body states.

\newpage
\begin{table}[tbp]
    \caption{The convergence of the total energies $E$ and expectation values of the 
             two-particle delta-functions for the ground (bound) $S(L = 0)-$state of 
             the $\overline{p}dt$ ion (in proton-atomic units). $N$ is the total number
             used in calculations.}
      \begin{center}
      \begin{tabular}{| c | c | c | c | c |}
        \hline\hline
 $N$ & $E$ & $\langle \delta_{\overline{p}t} \rangle$ &  
             $\langle \delta_{\overline{p}d} \rangle$ & $\langle \delta_{dt} \rangle$ \\  
      \hline
  3300 & -0.38119 08996 43549 79944 & 1.035725171596$\cdot 10^{-1}$ & 
          2.65020902768$\cdot 10^{-2}$ & 1.97194857$\cdot 10^{-4}$ \\

  3500 & -0.38119 08996 43549 80319 & 1.035725171777$\cdot 10^{-1}$ & 
          2.65020902581$\cdot 10^{-2}$ & 1.97194865$\cdot 10^{-4}$ \\
         
  3700 & -0.38119 08996 43549 80574 & 1.035725171405$\cdot 10^{-1}$ &
          2.65020903462$\cdot 10^{-2}$ & 1.97194853$\cdot 10^{-4}$ \\

  3840 & -0.38119 08996 43549 80725 & 1.035725171481$\cdot 10^{-1}$ &
          2.65020902513$\cdot 10^{-2}$ & 1.97194857$\cdot 10^{-4}$ \\
       \hline\hline
   \end{tabular}
   \end{center}
   \end{table}
\begin{table}[tbp]
    \caption{Some bound state properties of the ground (bound) $S(L = 0)-$state of 
             the $\overline{p}dt$ ion (in proton-atomic units). The particle 1 means
             deuteron, 2 desinates triton, while 3 stands for the antiproton.}
      \begin{center}
      \begin{tabular}{| c | c | c | c | c |}
        \hline\hline
 $\langle X \rangle$ & $(ij)$ = (32) & $(ij)$ = (31) & $(ij)$ = (21) \\ 
      \hline
 $\langle r^{-1}_{ij} \rangle$ & 0.6452584980114 & 0.3265829532244 & 0.20945965194864 \\

 $\langle (r_{ik} r_{jk})^{-1} \rangle$ & 0.090672197685358 & 0.13833503600526 & 0.17327231991926 \\

 $\langle r_{ik}^{-2} \rangle$ & 0.894094641294119 & 0.289502249538164 & 0.0635359440628808 \\
       \hline

 $\langle r_{ij} \rangle$      & 2.535959239261 & 5.822464028824 & 6.5332990137423 \\

 $\langle r^2_{ij} \rangle$    & 9.369595134746  & 52.35601818235 & 57.621835007682 \\

 $\langle r^3_{ij} \rangle$    & 46.65710362569 & 645.7495736644 & 676.62342378591 \\

 $\langle r^4_{ij} \rangle$    & 294.4990484287 & 10212.70615551 & 10290.298340574 \\
          \hline
 $\tau_{jk}$                   & 0.76584205188631 & 0.34062934917079 & 0.0766784190714  \\

 $F^{(a)}$                     & 0.084729933577141 & 0.05594048174487 & 1.217745967$\cdot 10^{-4}$ \\

 $\nu_{ij}$                    & -0.7496066938104 & -0.66655644341827 & 1.1986352684990 \\

 $\nu^{(b)}_{ij}$              & -0.7496066901228 & -0.66655635263091 & 1.1986366724385 \\
       \hline\hline
 $\langle p^2_{i} \rangle$     & 0.069022844172409 & 0.22440893908996 & 0.27170237010133 \\

 $\langle {\bf p}_{i} \cdot {\bf p}_{j} \rangle$   & 0.006716068397607 & 0.01060650083845 & 0.012940402539131 \\

 $\langle {\bf r}_{ik} \cdot {\bf r}_{jk} \rangle$ & 50.30412902764317 & 7.31770598003839 & 2.05188915470772 \\
        \hline
 $\langle r_{ik}^{-3} \rangle_R$ & -0.4103977567344 & -1.8132102882933 & 0.02747944473455 \\
      \hline\hline
   \end{tabular}
   \end{center}
${}^{(a)}$The values of $f, \langle (r_{12} r_{13} r_{23})^{-1} \rangle$ and $\langle \delta_{123} \rangle$, 
respectively. \\
${}^{(b)}$The exact values of the two-body cusps.  
   \end{table}
\begin{table}[tbp]
    \caption{The levels of hyperfine structure $\epsilon$ and hyperfine structure splittings 
             $\Delta$ of the ground bound $S(L = 0)-$state of the $\overline{p}dt$ ion (in $MHz$).}
      \begin{center}
      \begin{tabular}{| c | c | c |}
        \hline\hline
  & $\epsilon_{J}$ &  $\Delta$  \\ 
          \hline
$\overline{p}dt(J = 2)$ &  2.3642257710$\cdot 10^9$ & -----  \\ 

$\overline{p}dt(J = 1)$ &  2.2529733441$\cdot 10^9$ &  1.112524269$\cdot 10^8$ \\ 

$\overline{p}dt(J = 0)$ & -4.6001817067$\cdot 10^9$ &  6.853155051$\cdot 10^9$ \\

$\overline{p}dt(J = 1)$ & -4.6599557269$\cdot 10^9$ &  5.977402022$\cdot 10^7$ \\
      \hline\hline
   \end{tabular}
   \end{center}
   \end{table}


\begin{thebibliography}{10}

\bibitem{FrBi} D.M. Bishop, A.M. Frolov and V.H. Smith, Jr., Phys. Rev. A {\bf 51}, 3636 (1995).

\bibitem{FrTa} A.M. Frolov and A.J. Thakkar, Phys. Rev. A {\bf 46}, 4418 (1992).

\bibitem{Hu} E. Hu, Y.A. Asamo, M.Y. Chen et al, Nucl. Phys. A,  {\bf 254}, 403 (1975).

\bibitem{LLQ}L.D. Landau and E.M. Lifshitz, {\it Quantum Mechanics. 
Non-Relativistic Theory}, 3rd. edn. (Oxford, England, Pergamonn Press (1977)). 

\bibitem{Schmidt} K.M. Schmidt, Proc. R. Soc. (London), A {\bf 448}, 2829 (2002).

\bibitem{Fro2001} A.M. Frolov, Phys. Rev. E {\bf 64}, 036704 (2001); ibid, {\bf 74}, 027702 
(2006).

\bibitem{Fock} V.A. Fock, Izv. AN SSSR (ser. Fiz.) {\bf 18}, 161 (1954) [in Russian].

\bibitem{Fro07} A.M. Frolov, J. Phys. A {\bf 40}, 6175 (2007).

\bibitem{Barel} C. Amsler, Rev. Mod. Phys. {\bf 70}, 1293 (1998).

\bibitem{2009} M. Corradini, M. Hori, M. Leavy et al, Hyperfine Interactions {\bf 194}, 305 (2009).

\bibitem{1997} J. Carbonell, K.V. Protasov and A Zenoni,  Phys. Lett. B {\bf 397}, 345 (1997).

\bibitem{1999} A. Zenoni, A. Bianconi, F. Bocci et al,  Phys. Lett. B {\bf 461}, 405 (1999).

\bibitem{2001} A. Trzci\'{n}ska, J. Jastrzebski, P. Lubi\'{n}ski et al,  Phys. Rev. Lett. {\bf 87}, 
082501 (2001).

\bibitem{CRC} \textit{CRC Handbook of Chemistry and Physics}, 85th Edition,
Ed. D.R. Lide, (CRC Press, Inc., Boca Raton, Florida, 2004).

\bibitem{Babik} V.V. Babikov, {\it Method of Phase Functions in Quantum 
Mechanics}, (Nauka, Moscow, (1976)) [in Russian].

\end{thebibliography}
\end{document}